\documentclass[conference,10pt]{IEEEtran}

\usepackage{framed}

\usepackage{algorithm}
\usepackage{algpseudocode}
\usepackage{bm}
\usepackage{stackrel}
\usepackage{subfigure}
\usepackage{epsf}
\usepackage{amsmath,amssymb}
\usepackage{graphicx}
\usepackage{color}
\usepackage{cite}
\usepackage{multirow,tabularx}
\usepackage{ifthen}
\usepackage{epstopdf}
\input{epstopdf.sty}
\usepackage{times}

\input{epsf.sty}

\makeatletter
\def\BState{\State\hskip-\ALG@thistlm}
\makeatother

\newcommand{\hide}[1]{\ifthenelse{\boolean{false}}{#1}{}}

\newtheorem{theorem}{{\bf Theorem}}

\newtheorem{lemma}{{\bf Lemma}}

\newcommand{\qed}{\nobreak \ifvmode \relax \else
      \ifdim\lastskip<1.5em \hskip-\lastskip
      \hskip1.5em plus0em minus0.5em \fi \nobreak
      \vrule height0.75em width0.5em depth0.25em\fi}

\newcommand{\beq}{\begin{equation}}
\newcommand{\eeq}{\end{equation}}
\newcommand{\barr}{\begin{array}}
\newcommand{\earr}{\end{array}}

\newcommand{\benum}{\begin{enumerate}}
\newcommand{\eenum}{\end{enumerate}}

\newcommand{\bit}{\begin{itemize}}
\newcommand{\eit}{\end{itemize}}

\newcommand{\bc}{\begin{center}}
\newcommand{\ec}{\end{center}}

\newcommand{\bdes}{\begin{description}}
\newcommand{\edes}{\end{description}}

\newcommand{\bfig}{\begin{figure}}
\newcommand{\efig}{\end{figure}}

\newcommand{\bemq}{\begin{quote} \begin{em}}
\newcommand{\eemq}{\end{em} \end{quote}}

\newcommand{\bmp}{\begin{minipage}}
\newcommand{\emp}{\end{minipage}}

\newcommand{\brac}[1]{\left({#1}\right)}

\newcommand{\EX}[1]{\mathbb{E}\left[{#1}\right]} 

\newcommand{\prob}[1]{\text{Pr}\brac{#1}}

\newcommand{\bsp}{\begin{slide*}}
\newcommand{\esp}{\end{slide*}}
\newcommand{\bsl}{\begin{slide}}
\newcommand{\esl}{\end{slide}}

\newcommand{\blem}{\begin{lemma}}
\newcommand{\elem}{\end{lemma}}
\newcommand{\bthm}{\begin{theorem}}
\newcommand{\ethm}{\end{theorem}}

\newcommand{\pr}[1]{\mathbf{P}\left[ #1 \right]}

\IEEEoverridecommandlockouts

\begin{document}

\title{Distributed Scheduling Algorithms for Optimizing Information Freshness in Wireless Networks}

\author{Rajat Talak, Sertac Karaman, and Eytan Modiano
\thanks{The authors are with the Laboratory for Information and Decision Systems (LIDS) at the Massachusetts Institute of Technology (MIT), Cambridge, MA. {\tt \{talak, sertac, modiano\}@mit.edu} }
\thanks{Submitted to SPAWC 2018.}
\thanks{This work was supported by NSF Grants AST-1547331, CNS-1713725, and CNS-1701964, and by Army Research Office (ARO) grant number  W911NF-17-1-0508.}
}


\maketitle

\begin{abstract}
Age of Information (AoI), measures the time elapsed since the last received information packet was generated at the source. We consider the problem of AoI minimization for single-hop flows in a wireless network, under pairwise interference constraints and time varying channel. We consider simple, yet broad, class of distributed scheduling policies, in which a transmission is attempted over each link with a certain attempt probability. We obtain an interesting relation between the optimal attempt probability and the optimal AoI of the link, and its neighboring links. We then show that the optimal attempt probabilities can be computed by solving a convex optimization problem, which can be done distributively. 
\end{abstract}

\section{Introduction}
\label{sec:intro}
Timely delivery of information updates is gaining increasing relevance with the advent of technologies such as autonomous flying vehicles, internet of things, and other cyber physical systems. In autonomous flying vehicles, for example, timely exchange of position, velocity, and other control information can improve network safety~\cite{FANETs2014, talakCDC16}.

Packet delay, the traditionally used measure, accounts for the average time taken for a packet to reach the destination node, but does not measure the `information lag' at the destination node, especially when the information update generation can be controlled~\cite{2012Infocom_KaulYates, talak17_StGenIC_Mobihoc, talak17_allerton}. For example, if two packets carry the same information, and one arrives early at the destination, then there is no need in accounting for the delay in reception of the second packet.

Age of information (AoI), a newly proposed metric~\cite{2012Infocom_KaulYates, 2011SeCON_Kaul}, is defined to be the time elapsed since the last received information update was generated at the source node. AoI, upon reception of a new update packet drops to the time elapsed since generation of the packet, and grows linearly otherwise. AoI, therefore, more accurately captures the `information lag' at the destination node.


AoI has been analyzed for various queueing models~\cite{2012Infocom_KaulYates, 2015ISIT_LongBoEM, 2013ISIT_KamKomEp, 2012CISS_KaulYates, 2016ISIT_Najm, 2014ISIT_CostaEp}, but very little attention has been paid to AoI minimization in wireless networks.
A problem of scheduling finitely many update packets under physical interference constraints was shown to be NP-hard in~\cite{2016Ep_WiOpt}. Age for a broadcast network, where only a single link can be activated at any time, was studied in~\cite{2016allerton_IgorAge, 2017ISIT_YuPin, Igor18_infocom}. Some preliminary analysis of age for a slotted ALOHA like random access was done in~\cite{2017X_KaulYates_AoI_ALOHA}, while multi-hop networks have been considered in~\cite{BedewyISIT17_LIFO_opt, talak17_allerton}. More recent work includes~\cite{talak17_StGenIC_Mobihoc, talak17_ISIT, talak17_WiOpt}, which propose centralized scheduling policies for AoI minimization, under general interference constraints and time varying channels.

In a wireless network, it is not always possible to implement centralized schemes. In this paper, we consider a simple, yet broad, class of distributed scheduling policies for wireless networks, in which each link attempts transmission with a certain probability. We consider time varying channel and a pairwise interference model, in which each link interferes with a specified set of links in the network.

We first derive an interesting relation between the optimal attempt probability, and the optimal age of the given link and the interfering links.
We then characterize the optimal distributed policy $\pi_D$, as a solution to a convex optimization problem. We further show that this convex problem can be solved distributively.

\section{System Model}
\label{sec:model}
We model a wireless communication network as a graph $G = (V,E)$, where $V$ is the set of nodes and $E$ is the set of directed communication links between the nodes. Each link $e \in E$ is associated with a source-destination pair. Time is slotted, with the duration of each slot set equal to the time it takes to transmit a single update packet. For simplicity, we normalize the slot duration to unity.

Transmission of update packets cannot occur simultaneously over all links due to wireless interference constraints~\cite{ak_winet}. For each link $e \in E$, there is a subset of links $N_{e} \subset E$, that interfere with link $e$, i.e., if link $e$ and a link $l \in N_e$ attempt simultaneous transmission, then the transmission over link $e$ fails due to interference. We call this the \emph{pairwise interference model}. Popular models such as 1-hop and 2-hop interference models~\cite{ak_winet} are special cases of this model.

Channel uncertainty can also lead to failure in reception of update packets. We consider a two state channel process $S_{e}(t) \in \{0, 1\}$ for all time $t$ and links $e$. Channel state $S_{e}(t) = 1$ implies that link $e$'s channel is ON at time $t$, and a non-interfering transmission over $e$ is successfully received. When $S_{e}(t) = 0$, even a non-interfering transmission over link $e$ fails due to bad channel conditions.
We assume the channel process $\{ S_{e}(t) \}_{t, e}$ to be independent and identically distributed (i.i.d.) across time $t$, and only independent across links $e$. We let $\gamma_e = \prob{S_{e}(t) = 1}$ denote the channel success probability for link $e$. We consider the case when the channel statistics, namely $\gamma_e$ for all $e \in E$, are known but the current channel state $S_{e}(t)$ cannot be observed.

We use $U_{e}(t)$ to denote scheduling decisions at time $t$, for each link $e$. $U_{e}(t) = 1$ when link $e$ attempts transmissions in slot $t$, and $U_{e}(t) = 0$ otherwise. Thus, a successful transmission occurs over link $e$ if and only if a transmission is attempted on link $e$, the link $e$'s channel is ON, and no transmission is attempted over links $e' \in N_{e}$. This is equivalent to
\begin{equation}
\hat{U}_{e}(t) \triangleq S_{e}(t)U_{e}(t)\prod_{e' \in N_{e}}\left( 1 - U_{e'}(t)\right) = 1.
\end{equation}

\subsection{Age of Information}
AoI of a network is a function of AoI of each source-destination pair. Thus, we define age $A_{e}(t)$ for each link, to be the time since the last received update was generated.
We assume source nodes to be \emph{active sources}, i.e., it can generates a fresh update packet before every transmission. 

When all sources are active, the age $A_{e}(t)$, is equal to the time elapsed since the last successful transmission over $e$. Thus, the evolution of age $A_{e}(t)$ can be written as
\begin{equation}\label{eq:age_evol}
A_{e}(t+1) = \left\{ \begin{array}{cc}
                       A_{e}(t) + 1 &~\text{if}~\hat{U}_{e}(t) = 0 \\
                       1 &~\text{if}~\hat{U}_{e}(t) = 1
                     \end{array}\right.,
\end{equation}
for all $e \in E$.
We define average age for link $e$ to be
\begin{equation}\label{eq:ave_age}
A_{e}^{\text{ave}} = \limsup_{T \rightarrow \infty} \frac{1}{T}\sum_{t=1}^{T}A_{e}(t),
\end{equation}
whenever the limit exists. We define the peak age to be the average of all the age peaks:
\begin{equation}\label{eq:peak_age}
A_{e}^{\text{p}} = \limsup_{T \rightarrow \infty} \frac{1}{\mathcal{N}_{e}(T)}\sum_{i=1}^{\mathcal{N}_{e}(T)}A_{e}(T_{e}(i)),
\end{equation}
where $T_{e}(i)$ denote the $i$th instance of successful transmission over link $e$, i.e., the $i$th instance when $\hat{U}_{e}(t) = 1$, and $\mathcal{N}_{e}(T)$ is the number of successful transmissions over link $e$ until time $T$. We define average and peak AoI of the network to be a weighted sum of link AoI:
\begin{equation}\label{eq:aoi}
A^{\text{ave}} = \sum_{e \in E} w_e A_{e}^{\text{ave}},~~~\text{and}~~~A^{\text{p}} = \sum_{e \in E} w_e A_{e}^{\text{p}}.
\end{equation}

Our objective is to design distributed scheduling policies that minimize the network average and peak age.

\subsection{Distributed Stationary Policies}
We focus our attention on policies in which scheduling decisions are made by the source of each link $e$, and not by a centralized scheduler. In particular, we consider policies in which
each link $e$ attempts transmission with probability $p_e > 0$, independent across links and slots. We refer to these policies as the \emph{distributed stationary policies}.

The probability that a non-interfering transmission is attempted over link $e$ is given by $f_e = p_e \prod_{e' \in N_e} (1 - p_{e'})$. We will refer to $f_e$ as the \emph{link activation frequency} of link $e$, and use $\mathbf{f} = \left( f_e \right)_{e \in E}$ to denote the link activation frequency vector. The space of all link activation frequencies $\mathbf{f}$, attainable using  distributed stationary policies, is given by
\begin{multline}\label{eq:FD}
\mathcal{F}_D = \biggl\{ \mathbf{f} \in \mathbb{R}^{|E|}~|~f_e = p_e\prod_{e' \in N_e}(1 - p_{e'}) \\
\text{and}~0 \leq p_e \leq 1~\forall~e \in E \biggr\}.
\end{multline}

\begin{figure}
  \centering
  \includegraphics[width=0.65\linewidth]{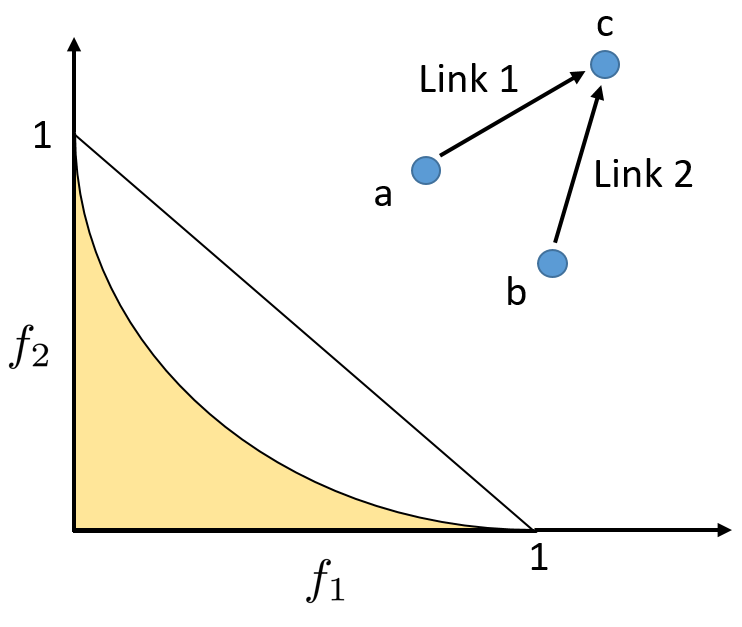} 
  \caption{Example of two interfering links, $1$ and $2$. The yellow region is $\mathcal{F}_D$. Also shown is the set of feasible link activation frequencies for centralized schedule: $\mathcal{F} = \{ (f_1, f_2)~|~f_1 + f_2 \leq 1~\text{and}~f_i \geq 0\}$.}\label{fig:example}
\end{figure}
Figure~\ref{fig:example} shows the set of feasible link activation frequencies for distributed policies for the two interfering link example.  Also, shown is the set of all link activation frequencies $\mathcal{F}$ attainable using a centralized scheduler. It can be seen that the set $\mathcal{F}_{D}$ is non-convex, and strictly smaller than $\mathcal{F}$. The gap between these two sets indicates the price one has to pay in resorting to distributed scheduling.

\section{Optimal Policy}
\label{sec:dist}
In this section, we characterize an age optimal policy, and propose an implementation that requires only local information. We first show that for any distributed stationary policy, the peak and average age are equal, and can be written as a simple convex function in link activation frequencies $f_e$.

\begin{framed}
\begin{lemma}
\label{lem:peak_ave_equal}
For any distributed stationary policy, the average and peak age is given by
\begin{equation}
A^{\text{ave}} = A^{\text{p}} = \sum_{e \in E}\frac{w_e}{\gamma_e f_e},
\end{equation}
where $f_e = p_e \prod_{e' \in N_{e}} (1 - p_{e'})$ is the link activation frequency of link $e$.
\end{lemma}
\end{framed}
\begin{IEEEproof}
A similar result was proved in~\cite{talak17_StGenIC_Mobihoc} which considered a class of centralized policies, that are not necessarily stationary. Here, we provide a proof for distributed stationary policies.

Consider a link $e$. Then, under any distributed stationary policy, the event that link $e$ is successfully activated in slot $t$ is independent and identically distributed across slots, with success probability of $\gamma_e f_e$. Thus, the time since last activation is geometrically distributed with mean $\frac{1}{\gamma_e f_e}$. It turns out that the peak and average age for link $e$ are indeed equal to this quantity, i.e.,
\begin{equation}
A^{\text{ave}}_{e} = A^{\text{p}}_{e} = \frac{1}{\gamma_e f_e}, \nonumber
\end{equation}
and therefore, the network age is $\sum_{e \in E} \frac{w_e}{\gamma_e f_e}$.
The detailed proof is given in Appendix~\ref{pf:lem:peak_ave_equal}.
\end{IEEEproof}

An immediate consequence of Lemma~\ref{lem:peak_ave_equal} is that both peak and average age minimization problems, over the space of all distributed stationary policies, can be written as
\begin{equation}
\label{eq:main_problem}
\begin{aligned}
& \underset{\mathbf{f} \in \mathcal{F}_{D}}{\text{Minimize}}
& & \sum_{e \in E} \frac{w_e}{\gamma_e f_e}
\end{aligned}
\end{equation}
Notice that, although the objective function is convex, the constraint set $\mathcal{F}_{D}$ is non-convex, see Figure~\ref{fig:example}. Furthermore, obtaining optimal link activation frequencies $\mathbf{f}^{\ast}$ does not suffice as the distributed policy is characterized by the attempt  probabilities $p_{e}$.
The optimal link attempt probabilities $\mathbf{p}^{\ast}$, that solve~\eqref{eq:opt_peak_age_problem_dist}, yield a distributed stationary policy, that is both peak and average age optimal. We call this policy $\pi_{D}$.

The following result characterizes the optimal link attempt probabilities $\mathbf{p}^{\ast}$ in terms of the optimal link age $A^{\ast}_{e}$.
\begin{framed}
\begin{theorem}
\label{thm:dist_opt}
The attempt probabilities $\mathbf{p}^{\ast} = \left(p^{\ast}_e| e \in E\right)$, that solve~\eqref{eq:main_problem}, are given by
\begin{equation}
\label{eq:fp}
p_{e}^{\ast} = \frac{ w_e A^{\ast}_{e}}{ w_e A^{\ast}_{e} + \sum_{e': e \in N_{e'}} w_{e'} A^{\ast}_{e'}},
\end{equation}
for all $e \in E$, where $A^{\ast}_{e} = \left[ \gamma_e p^{\ast}_{e} \prod_{e' \in N_e}\left( 1 - p_{e'}^{\ast}\right) \right]^{-1}$ is the optimal peak/average age of link $e$.
\end{theorem}
\end{framed}
\begin{IEEEproof}
Substituting $\mathcal{F}_{D}$, from~\eqref{eq:FD}, in~\eqref{eq:main_problem} we get
\begin{align}\label{eq:opt_peak_age_problem_dist}
\begin{aligned}
& \underset{\mathbf{p} \in [0,1]^{|E|}, \mathbf{f} \in \mathbb{R}^{|E|}}{\text{Minimize}}
& & \sum_{e \in E} \frac{w_e}{\gamma_e f_e} \\
& \text{subject to}
& & f_e = p_e \prod_{e' \in N_e}\left( 1 - p_{e'}\right)~\forall~e \in E \\
&&& f_e \geq 0~\text{for all}~e \in E
\end{aligned}
\end{align}
Now, substituting $q_{e} = 1 - p_{e}$, the optimization problem~\eqref{eq:opt_peak_age_problem_dist} reduces to
\begin{align}
\begin{aligned}
\label{eq:dist_opt}
&\underset{\mathbf{p} \geq 0, \mathbf{q} \geq 0}{\text{Minimize}} & &\sum_{e \in E} \frac{w_e}{\gamma_e p_e\prod_{e' \in N_e} q_{e'}} \\
&\text{subject to} & &p_{e} + q_{e} \leq 1 ~\forall e \in E
\end{aligned}
\end{align}
This is a convex program in standard form, and can be solved using KKT conditions to obtain Theorem~\ref{thm:dist_opt}. The details are given in  Appendix~\ref{pf:thm:dist_opt}.
\end{IEEEproof}


Preliminary results on age optimization for a ALOHA type network model were presented in~\cite{2017X_KaulYates_AoI_ALOHA} for a network in which only one link can be activated at any given time. As a heuristic, it was argued in~\cite{2017X_KaulYates_AoI_ALOHA} that the attempt rates should be
\begin{equation}\label{eq:local6}
p_e = \frac{ 1/\sqrt{\gamma_e} }{\sum_{e' \in E} 1/\sqrt{\gamma_{e'}} }.
\end{equation}
However, Theorem~\ref{thm:dist_opt} shows that the attempt probability
\begin{equation}
\label{eq:p_age}
p^{\ast}_e = \frac{ A^{\ast}_{e} }{\sum_{e' \in E} A^{\ast}_{e'} },
\end{equation}
is both peak and average age optimal. This gives us a more precise answer to the question posed in~\cite{2017X_KaulYates_AoI_ALOHA}. Our result, however, holds under the more general pairwise interference constraints, and non-equal weights $w_e > 0$.

\subsection{Distributed Computation of $\pi_D$}
Although, Theorem~\ref{thm:dist_opt} gives a characterization of the optimal attempt probabilities, in terms of the optimal link age $A^{\ast}_{e}$, it does not provide for a simple method to computation.

We now provide for an alternate characterization of the network age minimization problem~\eqref{eq:main_problem}. In particular, we show that $\mathbf{p}^{\ast}$ can be obtained by a simpler convex optimization problem, than the one in~\eqref{eq:dist_opt}. We will also see that this simpler problem is akin to distributed computation of the policy $\pi_D$.
\begin{framed}
\begin{theorem}
\label{thm:dual}
The age minimization problem~\eqref{eq:main_problem} is equivalent to the convex optimization problem:
\begin{align}
\label{eq:dual}
\begin{aligned}
& \underset{\lambda_{e} \geq 0}{\text{Maximize}}
    & & \!\!\!\! \sum_{e \in E}\left(\lambda_e +\!\!\!\!\! \sum_{e': e \in N_{e'}}\!\!\! \lambda_{e'}\right) H\!\left(\! \frac{\lambda_e}{\lambda_e + \sum_{e': e \in N_{e'}} \!\! \lambda_{e'}} \! \right) \\
    &&&+\sum_{e \in E} \lambda_e  \left[ 1 + \log\left( \frac{w_e}{\lambda_e}\right) \right] \triangleq G(\bm{\lambda}),
\end{aligned}
\end{align}
where $H\left( p \right) = p \log\left( \frac{1}{p}\right) + \left(1 - p\right)\log\left( \frac{1}{1-p}\right)$ is the entropy function. If $\bm{\lambda}^{\ast}$ is the optimal solution to this problem, then the optimal attempt probabilities $\mathbf{p}^{\ast}$ are given by
\begin{equation}
\label{eq:p_lambda}
p_{e}^{\ast} = \frac{\lambda_{e}^{\ast}}{\lambda_{e}^{\ast} + \sum_{e': e \in N_{e'}} \lambda_{e'}^{\ast}},
\end{equation}
for all $e \in E$.
\end{theorem}
\end{framed}
\begin{IEEEproof}
See Appendix~\ref{pf:thm:dual}.
\end{IEEEproof}

We first note that the variable $\lambda_e$ is a proxy for the weighted age $w_e A_{e}$ of link $e$, and the solution $\lambda^{\ast}_{e}$ corresponds to optimal weighted age $w_e A^{\ast}_{e}$. This can also be intuited from~\eqref{eq:p_lambda}, which is same as~\eqref{eq:fp}.

The objective function $G(\bm{\lambda})$ in~\eqref{eq:dual} is a concave function; since the problem is convex. Therefore, a simple projected gradient ascent is guaranteed to converge to the optimal $\bm{\lambda}^{\ast}$~\cite{boyd}. The gradient of the function $G\left( \bm{\lambda} \right)$ is given by
\begin{equation}
\nonumber
\frac{\partial G\left( \bm{\lambda} \right) }{\partial \lambda_e} = \log\left( \frac{w_e}{\lambda_{e}} \right) + \log\left( 1\! +\! \frac{\theta_{e}}{\lambda_{e}}\right) + \!\!\!\!\!\sum_{e': e \in \in N_{e'}} \!\!\!\!\! \log\left( 1 + \frac{\lambda_{e'}}{\theta_{e'}}\right),
\end{equation}
for all $e \in E$, where $\theta_e = \sum_{e': e \in N_{e'}} \lambda_{e'}$.
Note that the gradient $\frac{\partial G\left( \bm{\lambda} \right) }{\partial \lambda_e}$ depends only on $\lambda_e$ and $\lambda_{e'}$, for $e'$ that are `extended neighbors' of $e$. This space dependence of $\frac{\partial G\left( \bm{\lambda} \right) }{\partial \lambda_e}$ on $\bm{\lambda}$ makes the projected gradient descent algorithm akin to distributed implementation.

We give the projected gradient descent algorithm for distributively computing policy $\pi_D$ in Algorithm~\ref{algo:dist_policy}. In it, we divide time into frames, each frame of $F \geq 1$ slots. For simplicity, we assume symmetric interference, i.e., $e$ interferes with $e'$ if and only if $e'$ interferes with $e$ which is same as the condition $N_{e} = \{ e' \in E~|~e \in N_{e'} \}$ for all links $e$.

Link $e$, sets its attempt probability to $p_{e}(m)$, in frame $m \geq 1$. The source of link $e$ tracks and updates two variables, namely, $\lambda_{e}(m)$ and $\theta_{e}(m)$. We use $\Pi_{\epsilon}$ to denote the projection on the set $[\epsilon, +\infty)$, for a $\epsilon > 0$. As seen in Steps~5 and~7 in Algorithm~\ref{algo:dist_policy}, variables $\lambda_{e}(m)$ and $\theta_{e}(m)$ need to be exchanged only between the neighboring links.

\begin{algorithm}
\caption{Distributed Computation of Policy $\pi_D$}
\label{algo:dist_policy}
\begin{algorithmic}[1]
\BState $p_{e}(m)$: attempt probability of link $e$, in frame $m$
\BState $\lambda_{e}(m), \theta_{e}(m)$: dual variables, in frame $m$
\BState \textbf{Start}: Set $\lambda_{e}(0) = 1$ and $\theta_{e}(0) = |N_{e}|$ for all $e \in E$. Set $p_{e}(0) = 1/2$ and $m = 0$.
\For  {each frame $m$}
\State \textbf{Send}: $\theta_{e}(m)$ to all $e' \in N_{e}$
\State Compute $\lambda_{e}(m+1)$:
        \begin{multline}
        \nonumber
        ~~~\lambda_{e}(m+1) \gets \Pi_{\epsilon}\Bigg[\lambda_{e}(m) + \eta_{m} \Bigg\{ \log\left( \frac{w_e}{\lambda_{e}(m)} \right) \\
        + \log\left( 1 + \frac{\theta_{e}(m)}{\lambda_{e}(m)}\right)
        +\sum_{e' \in N_{e}} \log\left( 1 + \frac{\lambda_{e'}(m)}{\theta_{e'}(m)}\right) \Bigg\}\Bigg]
        \end{multline}
\State \textbf{Send}: $\lambda_{e}(m+1)$ to all $e' \in N_{e}$
\State Update $\theta_{e}(m+1)$:
        \begin{equation}
        \nonumber
        \theta_{e}(m+1) \gets \sum_{e' \in N_{e}} \lambda_{e'}(m+1)
        \end{equation}
\State Update $p_{e}(m+1)$:
        \begin{equation}
        \nonumber
        p_{e}(m+1) \gets \frac{\lambda_{e}(m+1)}{\lambda_{e}(m+1) + \theta_{e}(m+1)}
        \end{equation}
\State $m \gets m+1$
\EndFor
\end{algorithmic}
\end{algorithm}

%
%

%
%
%


\section{Conclusion}
\label{sec:conclusion}
We considered age minimization in a wireless network, with pairwise interference constraints, time varying channels, and single-hop flows. We considered a simple class of distributed scheduling policies, in which each link attempts transmission with probability $p_e$. We showed an interesting relationship between the optimal attempt probability for a link, and the optimal age of the link and it's neighboring links. We then showed that the optimal link attempt probabilities can be obtained by solving a convex optimization problem, which can be done distributively using the projected gradient ascent algorithm.

\bibliographystyle{ieeetr}

\appendix

\subsection{Proof of Lemma~\ref{lem:peak_ave_equal}}
\label{pf:lem:peak_ave_equal}
Consider a distributed stationary policy with link activation frequencies $f_e > 0$ for each $e \in E$. For a link $e$, let $T_{e}(i)$ be the $i$th instance when link $e$ was successfully activated, i.e., the $i$th instance when $\hat{U}_{e}(t) = 1$. Let $X_{e}(i) = T_{e}(i) - T_{e}(i-1)$ to be the $i$th inter-(successful) activation time for link $e$. Since all the processes involved, namely $S_{e}(t)$ and $U_{e}(t)$, are i.i.d. across time $t$, $X_{e}(i)$ is geometrically distributed with rate equal to $\pr{\hat{U}_{e}(t) = 1} = \gamma_e f_e$, i.e., $\pr{X_{e}(i) = k} = \gamma_e f_e \left( 1 - \gamma_e f_e\right)^{k-1}$,
for all $k \in \{1, 2, \ldots \}$.

We now compute the average age $A^{\text{ave}}_{e}$ for link $e$. Note that the age $A_{e}(t)$ over slots $t \in \{ T_{e}(i)+1, \ldots T_{e}(i+1)\}$ increases from $1$ to $X_{e}(i)$ in steps of $1$. Therefore, $\sum_{t=T_{e}(i)+1}^{T_{e}(i+1)} \!\!\!\! A_{e}(t) = \sum_{m=1}^{X_{e}(i)}m$
for all $i$. Then, using renewal theory~\cite{wolff}, we can derive the average age to be
\begin{align}
A^{\text{ave}}_{e} &= \limsup_{T \rightarrow \infty} \frac{1}{T}\sum_{t=1}^{T}A_{e}(t) = \lim_{N \rightarrow \infty} \frac{\sum_{i=1}^{N} \sum_{m=1}^{X_{e}(i)}m}{\sum_{i=1}^{m} X_{e}(i)}, \nonumber \\
&= \lim_{N \rightarrow \infty} \frac{\sum_{i=1}^{N}\frac{1}{2}X_{e}(i)\left(X_{e}(i)+1\right)}{\sum_{i=1}^{N}X_{e}(i)}, \nonumber \\
&= \frac{1}{2}\frac{\EX{X_{e}(i)\left( X_{e}(i) + 1\right)}}{\EX{X_{e}(i)}}~~\text{a.s.}.
\end{align}
Computing the moments of $X_{e}(i)$ we get $A^{\text{ave}}_{e} = \frac{1}{\gamma_e f_e}$ a.s..

For peak age, note that the $i$th peak is equal to the $i$th inter-(successful) activation time $X_{e}(i)$, i.e., $A_{e}\left( T_{e}(i)\right) = X_{e}(i)$. Therefore, we have
\begin{equation}
\label{eq:peak_age_formula}
A^{\text{p}}_{e} = \limsup_{T \rightarrow \infty} \frac{1}{\mathcal{N}(T)}\sum_{i=1}^{\mathcal{N}(t)}A_{e}\left( T_{e}(i)\right) = \lim_{N \rightarrow \infty} \frac{1}{N}\sum_{i=1}^{N}X_{e}(i), \nonumber
\end{equation}
which is equal to $\EX{X_{e}(1)} = \frac{1}{\gamma_e f_e}$ a.s.. This proves the result.

\subsection{Proof of Theorem~\ref{thm:dist_opt}}
\label{pf:thm:dist_opt}
Consider the Lagrangian dual
\begin{equation}
L(\mathbf{p}, \mathbf{q}, \bm{\lambda}) = \sum_{e \in E} \frac{w_e}{\gamma_e p_e \prod_{e' \in N_{e}} q_{e'}} + \sum_{e \in E} \lambda_{e}\left( p_e + q_e - 1\right).
\end{equation}
The problem~\eqref{eq:dist_opt} satisfies Slater's conditions, and thus, the KKT conditions are both necessary and sufficient. Setting partial derivatives of $L$ with respect to $p_e$ and $q_e$ to zero, we obtain
\begin{equation}
p_e\lambda_e = w_e A_{e},~~\text{and}~~q_e\lambda_e = \sum_{e': e \in N_{e'}} w_{e'} A_{e'},\label{eq:tt1}
\end{equation}
for all $e \in E$, where $A_{e} = \frac{1}{\gamma_e p_{e}\prod_{e' \in N_{e}}q_{e'}}$ is the age of link $e$. Equations~\eqref{eq:tt1} imply that $\lambda_e$ cannot be zero. By complementary slackness criteria we must have $p_e + q_e = 1$. This, with~\eqref{eq:tt1}, gives $p_e = \frac{w_e A_e}{w_e A_e + \sum_{e': e \in N_{e'}} w_{e'} A_{e'} }$
and $\lambda_e = w_e A_e + \sum_{e': e \in N_{e'}} w_{e'} A_{e'}$. This proves the result.

\subsection{Proof of Theorem~\ref{thm:dual}}
\label{pf:thm:dual}

The age minimization problem, given in~\eqref{eq:opt_peak_age_problem_dist}, can be written as by substituting $q_{e} = 1 - p_{e}$:
\begin{align}
\label{eq:opt_0}
\begin{aligned}
& \underset{\mathbf{f} \geq 0, \mathbf{p} \geq 0, \mathbf{q} \geq 0}{\text{Minimize}}
    & &  \sum_{e \in E} \frac{w_e}{\gamma_e f_{e}} \\
    & \text{subject to}
    & & f_{e} \leq p_{e} \prod_{e' \in N_e} q_{e'}~\forall~e \in E\\
    & & & p_e + q_e \leq 1~\forall~e \in E
\end{aligned}
\end{align}
Now, substituting $P_e = \log p_e$, $Q_e = \log q_e$, and $h_{e} = \log f_e$, the problem reduces to
\begin{align}
\label{eq:opt}
\begin{aligned}
& \underset{\mathbf{h}, \mathbf{P}, \mathbf{Q}}{\text{Minimize}}
    & & \sum_{e \in E} \frac{w_e}{\gamma_e}\exp\left\{ - h_{e}\right\}  \\
    & \text{subject to}
    & & h_{e} - P_e - \sum_{e' \in N_e} Q_{e'} \leq 0~\forall~e \in E\\
    & & & \log \left( \exp\left\{ P_e \right\} + \exp\left\{ Q_e \right\} \right) \leq 0~\forall~e \in E
\end{aligned}
\end{align}
This follows by first making the the substitution $p_e = e^{P_e}$, $q_e = e^{Q_e}$, and $f_e = e^{h_e}$, and then taking $\log$ in each of the constraints. This is a standard technique in geometric programming~\cite{boyd}. However, unlike geometric programming, we don't need to apply log function to the objective as it is already convex, and also separable in variables $h_e$.

The problem~\eqref{eq:opt} is convex~\cite{boyd}, and satisfies slater's conditions. As a consequence, the duality gap is zero, and the problem~\eqref{eq:opt} is equivalent to its Lagrangian dual problem.

Define the Lagrangian function of the optimization problem~\eqref{eq:opt} to be
\begin{multline}\label{eq:lagrangian}
L\left(\mathbf{h},\mathbf{P},\mathbf{Q},\bm{\lambda},\bm{\nu}\right) = \sum_{e \in E} \frac{w_e}{\gamma_e} e^{-h_{e}} + \sum_{e \in E} \nu_{e} \log \left( e^{P_e} + e^{Q_e}\right) \\
+ \sum_{e \in E} \lambda_e \big( h_e - P_e - \sum_{e' \in N_e} Q_{e'} \big),
\end{multline}
for $\bm{\lambda} \geq 0$ and $\bm{\nu} \geq 0$. The dual problem for~\eqref{eq:opt} is given by
\begin{align}
\label{eq:temp_dual}
\begin{aligned}
& \underset{\bm{\lambda} \geq 0, \bm{\nu} \geq 0}{\text{Maximize}} & & g\left(\bm{\lambda}, \bm{\nu} \right),
\end{aligned}
\end{align}
where $g\left(\bm{\lambda}, \bm{\nu} \right)$ is the dual objective function defines as
\begin{align}
\label{eq:x}
\begin{aligned}
g\left(\bm{\lambda}, \bm{\nu} \right) =& \underset{\mathbf{h}, \mathbf{P}, \mathbf{Q}}{\text{Minimize}} & & L\left(\mathbf{h},\mathbf{P},\mathbf{Q},\bm{\lambda},\bm{\nu}\right).
\end{aligned}
\end{align}
We will now show that the the optimization problem in Theorem~\ref{thm:dual} is in fact the dual problem~\eqref{eq:temp_dual}.

First note that the Lagrangian function  $L\left(\mathbf{h},\mathbf{P},\mathbf{Q},\bm{\lambda},\bm{\nu}\right)$ is strictly convex in $\mathbf{h}$, $\mathbf{P}$, and $\mathbf{Q}$~\cite{boyd}. Thus, first order conditions $\nabla_{\mathbf{h},\mathbf{P},\mathbf{Q}} L = 0$ yield the optimal  $\mathbf{h}$, $\mathbf{P}$, and $\mathbf{Q}$. Setting these partial derivatives to $0$, we get
\begin{equation}
\label{eq:condPQ}
\lambda_e =  \frac{ \nu_e e^{P_e}}{ e^{P_e} + e^{Q_e} }, \sum_{e': e \in N_{e'}}\!\!\!\!\! \lambda_{e'} =  \frac{\nu_e e^{Q_e}}{ e^{P_e} + e^{Q_e} },~\text{and}~\lambda_e = \frac{w_e}{\gamma_e} e^{-h_{e}},
\end{equation}
for all $e \in E$. Adding and dividing the first two equations in~\eqref{eq:condPQ} also yields
\begin{equation}\label{eq:conPQimp}
\nu_e = \lambda_e +\!\!\! \sum_{e': e \in N_{e'}}\!\!\!\!\! \lambda_{e'},~\text{and}~P_e - Q_e = \log\! \left( \!\frac{\lambda_e}{ \sum_{e' : e \in  N_{e'}}\! \lambda_{e'} }\!\right),
\end{equation}
respectively, for all $e$. Substituting~\eqref{eq:condPQ} and~\eqref{eq:conPQimp} in L, in~\eqref{eq:lagrangian}, we recover $g(\bm{\lambda}, \bm{\mu}) = G(\bm{\lambda})$, the objective function in~\eqref{eq:dual},
and therefore, the dual problem~\eqref{eq:temp_dual} is same as the optimization problem in Theorem~\ref{thm:dual}. This proves the first part.

We now show that if $\bm{\lambda}^{\ast}$ is the solution to this problem, then the optimal attempt probability is given by~\eqref{eq:p_lambda}. Let $\mathbf{f}^{\ast}$, $\mathbf{p}^{\ast}$, $\mathbf{q}^{\ast}$ be the solution to~\eqref{eq:opt_0}, $\mathbf{h}^{\ast}$, $\mathbf{P}^{\ast}$, $\mathbf{Q}^{\ast}$ be the solution to~\eqref{eq:opt}, and $\bm{\lambda}^{\ast}$ be the solution to the problem in Theorem~\ref{thm:dual}.

Note that, for each link $e \in E$, we must have $f^{\ast}_{e} = p^{\ast}_{e}\prod_{e' \in N_e}q^{\ast}_{e}$ and $p^{\ast}_{e} + q^{\ast}_{e} = 1$; as otherwise the objective function can be reduced by increasing $f^{\ast}_e$ and/or $p_{e}^{\ast}$. Since $p^{\ast}_e = e^{P^{\ast}_{e}}$ and $q_{e}^{\ast} = e^{Q_{e}^{\ast}}$, we have $e^{P^{\ast}_{e}} + e^{Q^{\ast}_{e}} = 1$. Substituting this in~\eqref{eq:condPQ},  and using~\eqref{eq:conPQimp}, we obtain
\begin{equation}
p^{\ast}_e = e^{P^{\ast}_{e}} = \frac{\lambda^{\ast}_{e}}{\lambda_{e}^{\ast} + \sum_{e': e \in N_{e'}} \lambda^{\ast}_{e'}},
\end{equation}
which proves the result.

\end{document}